\documentclass[12pt,fleqn,cite,epsfig]{article}
\usepackage{cite}
\usepackage[english]{babel}
\usepackage{empheq}
\usepackage{url}
\usepackage[utf8]{inputenc}
\usepackage[toc,page]{appendix}
\usepackage[autostyle]{csquotes}
\usepackage{slashed}
\usepackage{graphicx}
\usepackage{longtable}
\usepackage{tabularx}
\usepackage{amsmath}
\usepackage{nccmath}
\usepackage{cancel}
\usepackage{mathtools}
\usepackage{amsfonts}
\usepackage{amssymb}
\usepackage[paper=a4paper,
            marginparwidth=0in,     
            marginparsep=.05in,       
            margin=.65in,               
            includemp]{geometry}
\usepackage{hyperref}
\usepackage{cleveref}
\crefname{section}{§}{§§}
\Crefname{section}{Appendix}{Appendices}

\newcommand{\bse}{\begin{subequations}}
\newcommand{\ese}{\end{subequations}}
\numberwithin{equation}{section}
\title{\textbf{General Heat Kernel Coefficients for Massless Free Spin-3/2 Rarita-Schwinger Field}}

\author{Sudip Karan\footnote{\url{sudip.karaan@gmail.com}},  Shashank Kumar \footnote{\url{shashankshimla23@gmail.com}} and Binata Panda\footnote{\url{ binata@iitism.ac.in}}\\~\\
 \textit{Department of Applied Physics}\\
 \textit{ Indian Institute of Technology (Indian School of Mines)}\\\textit{Dhanbad, Jharkhand-826004, INDIA}}
\date{}
\begin{document}
\maketitle
\begin{abstract}
We review the general heat kernel 
method for the Dirac spinor field as an elementary example
in any arbitrary background.
We, then compute the first three Seeley-DeWitt coefficients for the massless free spin-3/2 Rarita-Schwinger field
without imposing any limitations on the background geometry.
\end{abstract}
\pagebreak                                                        
\section{Introduction}
Heat kernel is the fundamental solution to the heat equation for an elliptic partial 
differential operator on a specified domain with proper boundary conditions. 
 It turns out that the heat kernel is one of the most important tools
in spectral geometry, in 
differential geometry, and in global analysis more generally \cite{Gilkey:1995}.
Moreover, asymptotics of the heat kernel
are related to the eigenvalue asymptotics \cite{Gilkey:1973cc}. It is also a useful
tool to study the index theorem of Atiyah and
Singer \cite{Atiyah:1963aa,Atiyah:1973bb}.

Another motivation to study the heat kernel comes from physics. 
The important works by Fock \cite{Fock:1937dd} and Schwinger \cite{Schwinger:1951ee}
introduced the heat kernel to quantum theory. 
Again, DeWitt's covariant approach \cite{DeWitt:1965ff,DeWitt:1967gg,DeWitt:1967hh,DeWitt:1967ii} made 
heat kernel  one of the powerful 
tools for quantum field theory and quantum gravity 
(also review \cite{Avramidi:2002jj,Avramidi:2000kk,Vassilevich:2003ll}).
The heat kernel expansion 
is a very convenient tool for studying
one-loop divergences, anomalies, and various asymptotics of the
effective action. In recent years, the progress in theoretical physics, especially in string theory
 and related areas made special use of heat kernel coefficients to calculate 
the logarithmic corrections to Bekenstein-Hawking black hole entropy 
\cite{Charles:2015nn,Banerjee:2011oo,Banerjee:2011pp,Sen:2012qq,Sen:2012rr,Bhattacharyya:2012ss}.
These logarithmic corrections have great significance as 
they are aspects of the high energy theory, but one may compute them systematically in 
the low energy effective theory;
which involves computation of heat kernel coefficient 
(more precisely coefficient $a_4$ introduced in \cref{coefficients}) for 
the kinetic differential operator of the massless fields in the black hole background.

In recent years, Rarita–Schwinger field has become one of the active research topics 
in theoretical physics. Although, Rarita–Schwinger field plays dominating role in supergravity,
it has many other important applications. 
Analysis for massless Rarita–Schwinger field  
in covariant and the Coulomb-like gauges can be useful for perturbative interacting field theories 
\cite{Manoukian:2016ppp}. 
Also, recent investigations \cite{Maroto:2000aab,Nilles:2001aac,Kallosh:2000aad} 
suggest that spin-3/2 may have a crucial role to play in explaining the evolution of the universe.
It turns out that a detailed investigation of the heat kernel for the spin-3/2 Rarita–Schwinger field is
useful for  many research applications. These heat kernel coefficients for different fields were used to study 
quantum properties of different higher dimensional and dimensionally reduced supersymmetric theories 
\cite{Fradkin:1982abxy,Fradkin:1983acxz}.
The heat kernel coefficients for spin-3/2 field in general covariant gauge
are expressed in terms of the  heat kernel coefficients
for spin-1/2 field
following DeWitt's algorithm \cite{Endo:1995wz}.
However, in the present paper, we are going to represent
a general approach for calculating heat kernel coefficients
for massless free spin-3/2 Rarita-Schwinger field in any arbitrary background. 
We start with  reviewing the general heat kernel 
technique for some elementary examples of Dirac spinor field around an arbitrary background field configuration.

To illustrate the use of heat kernel in quantum field theory, 
let us consider the Gaussian form of the Euclidean path integral 
in one-loop approximation\footnote{One loop approximation involves the expansion of the Lagrangian $\mathcal{L}(\Phi)$ 
of the system up to quadratic order in fluctuations ($\phi$) around some classical background.}
\begin{ceqn}
\begin{equation}
 e^{-W} = \int [\mathcal{D}\phi] \thinspace e^{-\phi\Lambda\phi}= \frac{1}{\sqrt{\text{det}\Lambda}},
\end{equation}
where $\Lambda$ is a kinetic differential 
operator of background fields $\{\phi\}$ (schematic form
is shown in \cref{lambda}) and $W$ is the corresponding one-loop effective action. 
Also, one can express the one loop determinant det$\Lambda$ as a product over the spectrum of $\Lambda$,
\begin{equation}
\text{det}\Lambda = \prod_i \lambda_i,
\end{equation}
\begin{equation}\label{ld}
  W = \frac{1}{2}\thinspace \text{ln det} \thinspace \Lambda = \frac{1}{2}\sum_i \text{ln}\thinspace \lambda_i.
\end{equation}

We regularize the sum over eigenvalues ($\lambda_i's $) using the heat kernel method.
With the eigenfunctions of $\Lambda$ are given, 
one can modify the Green’s function\footnote{The Green’s function is 
subject to the boundary condition $K(x,x^\prime;s=0) = \delta(x-x^\prime)$,
a consequence of the orthonormality of the eigenfunctions $f_i(x)$.} as
\begin{equation}
K(x,x^\prime;s) = \sum_i e^{-s\lambda_i}f_i(x)f_i^*(x^\prime),
\end{equation}
which satisfies the heat equation
\begin{equation}
 (\partial_s + \Lambda_x)K(x,x^\prime;s) = 0;
\end{equation}
hence the reference to heat. 
The heat kernel parameter $s$ has units of length squared. One can avoid the dependence 
on the eigenfunctions by setting $x=x^\prime$, integrating over the manifold of interest, and using the normalization of the eigenfunctions $f_i(x)$.
The resulting function $D(s)$ will be referred as the \textit{heat kernel},
\begin{equation}\label{kernel}
D(s) = \int d^4x \sqrt{\text{det}\thinspace g_{\mu\nu}}\thinspace K(x,x;s) = 
\text{tr}\thinspace (e^{-s\Lambda}) = \sum_i e^{-s\lambda_i}.
\end{equation}
$D(s)$ is an object of interest for us because the one loop determinant \eqref{ld} is then expressed as
\begin{equation}\label{hk}
 W = -\frac{1}{2}\int_{\epsilon^2}^\infty ds\frac{D(s)}{s},
\end{equation}
where $\epsilon$ is a UV regulator with the dimension of length,
added by hand since the integral \eqref{hk} diverges at small $s$. 
$W$ is important to our work as it is the input to Sen’s quantum entropy 
function. Given the spectrum $\lambda_i$ of the kinetic operator $\Lambda$,
one can evaluate the heat kernel using \cref{kernel}. Once we get $D(s)$ in our 
hand, we can calculate the one loop effective action $W$ and 
thus one loop corrections to black hole entropy, given the knowledge of the spectrum 
of fields present around the black hole geometry.

The rest of this paper is organized as follows.
In \cref{method}, we first review the general heat kernel method 
\cite{Vassilevich:2003ll} for determining heat kernel coefficients 
for field fluctuations around any  arbitrary background, and
also list out general formulae for first three Seeley-DeWitt coefficients.
Next, we employ this method to the  different cases of Dirac spinor field as a warm up exercise.
We consider the cases of free and gauged Dirac spinor fields for both massless and massive cases.
Then, in \cref{sec3} we compute Seeley-DeWitt coefficients
for the spin-3/2 Rarita-Schwinger field following the 
same general approach and comment on the consistency of the results. In \cref{diss},
we end by summarizing the results.
\section{General heat kernel method}\label{method}

Let us start with a four-dimensional theory of gravity with an action having the form,
\begin{equation}
\mathcal{S} = \int d^4x \sqrt{\text{det}\thinspace g_{\mu\nu}} \thinspace \mathcal{L},
\end{equation}
where $g_{\mu\nu}$ is the background metric\footnote{The Euclidean signature of the metric is assumed in this paper; 
in our convention $\text{sig}(g_{\mu\nu}) = (++++)$.}
and $\mathcal{L}$ is the Lagrangian density.
Considering  upto quadratic order in fluctuations around a particular background, the action takes the  schematic form,
\begin{equation}
\mathcal{S} = \int d^4x \sqrt{\text{det}\thinspace g_{\mu\nu}}\thinspace M^{mp}\phi_p \Lambda_m^n\phi_n,
\end{equation}
where  $M$ is some matrix, \{$\phi_m$\} is the set of all the fluctuating fields,
$\Lambda$ is the kinetic differential operator acting on quadratic field fluctuations.

The present method demands $\Lambda$ to be both Hermitian and Laplace-type.
To make $\Lambda$ as Hermitian, we must restructure our action (up to a total derivative)
while for the later requirement, we choose $\Lambda$ to have the following Laplace-type form
\begin{equation}\label{lambda}
\Lambda_m^n = -(D^\mu D_\mu)I_m^n - (N^\mu D_\mu)_m^n-P_m^n,
\end{equation}
where $D_\mu$ is the ordinary covariant derivative with 
connections determined by the background metric;
$N$ and $P$ are arbitrary matrices constructed from the background fields;
and $I$  is the corresponding identity operator on the space of fields.

Using the definition $\mathcal{D}_\mu = D_\mu+\omega_\mu$ in \cref{lambda} we can rewrite $\Lambda$ as
\begin{equation}\label{compp}
\Lambda_m^n = -(\mathcal{D}^\mu\mathcal{D}_\mu)I_m^n-\left\lbrace P_m^n-(\omega^\mu\omega_\mu)_m^n-(D^\mu\omega_\mu)_m^n\right\rbrace-\left\lbrace (N^\mu-2\omega^\mu)D_\mu\right\rbrace_m^n.
\end{equation}
Finally, one can compare \cref{compp} with the standard form
\begin{equation}\label{comp}
\Lambda = -(\mathcal{D}^\mu\mathcal{D}_\mu)I-E,
\end{equation}
and identify the following useful matrices
\begin{equation}\label{geb}
\begin{gathered}
\omega_\mu = \frac{1}{2}N_\mu,\\
E = P -(\omega^\mu\omega_\mu)-(D^\mu\omega_\mu).
\end{gathered}
\end{equation}

Note that the term in parentheses $(D^\mu\omega_\mu)$  indicates the covariant derivative $D^\mu$ acting only on $\omega_\mu$.
The curvature associated with $\mathcal{D}_\mu$ is defined as
\begin{equation}\label{omega}
\begin{split}
\Omega_{\mu\nu} &\equiv [\mathcal{D}_\mu,\mathcal{D}_\nu]\\
&= [D_\mu,D_\nu]+D_{[\mu}\omega_{\nu]}+[\omega_\mu,\omega_\nu].
\end{split}
\end{equation}


\subsection{General formulae}\label{formulae}
The heat kernel corresponding to the differential operator
$\Lambda$ is defined in \cref{kernel}, and for small $s$ its perturbative expansion has
the following form \cite{Seeley:1966tt,Seeley:1969uu,DeWitt:1965ff,DeWitt:1967gg,DeWitt:1967hh,
DeWitt:1967ii,Duff:1977vv,Christensen:1979ww,Christensen:1980xx,Duff:1980yy,Birrel:1982zz,Gilkey:1984xy,Duff:2011yz,Vassilevich:2003ll}
\begin{equation}\label{sd}
\begin{split}
D(s) &\cong \int d^4x \sqrt{\text{det}\thinspace g_{\mu\nu}}\thinspace\sum_{n=0}^\infty s^{n-2}a_{2n}(x)\\
&= \int d^4x \sqrt{\text{det}\thinspace g_{\mu\nu}}\thinspace\left\lbrace\frac{1}{s^2}a_0(x) + \frac{1}{s}a_2(x) + a_4(x)+\cdots \right\rbrace,
\end{split}
\end{equation}

where the coefficients $a_{2n}(x)$ are known as \textit{Seeley-DeWitt coefficients}\footnote{In \cref{sd}, the terms $\int d^4x \sqrt{\text{det}\thinspace g_{\mu\nu}}\thinspace a_{2n}(x)$ are 
our desired \textit{heat kernel coefficients} and to obtain these coefficients,
we must choose a particular manifold to perform the integration. 
But our aim is to calculate generalized form heat kernel coefficients for spin-3/2 Rarita-Schwinger field. 
So, we focus on computation of Seeley-DeWitt coefficients instead.} 
\cite{Seeley:1966tt,Seeley:1969uu,DeWitt:1965ff,DeWitt:1967gg,DeWitt:1967hh,DeWitt:1967ii}. 
We want to express these coefficients in terms of local invariants obtained from background metric ($g_{\mu\nu}$),
Riemann tensor ($R_{\mu\nu\rho\sigma}$), Ricci tensor ($R_{\mu\nu}$), Ricci scalar ($R$), 
gauge field strength ($F_{\mu \nu}$) and their covariant derivatives. Once computed we can employ this general result 
for any arbitrary background field configuration.

With the definitions discussed in this section, the results for
the leading Seeley-DeWitt coefficients (see eqs. 4.26, 4.27 and 4.28 of \cite{Vassilevich:2003ll})
are summarized in the following equations\footnote{The formulae given in 
\cref{coefficients} are only valid for manifolds without boundary.}
\begin{equation}\label{coefficients}
\begin{gathered}
(4\pi)^2a_0(x) = \text{tr}(I),\\
(4\pi)^2a_2(x) = \frac{1}{6}\text{tr}(6E+RI),\\
(4\pi)^2a_4(x) = \frac{1}{360} \text{tr}\left\lbrace 60RE+180E^2 + 30\Omega_{\mu\nu}\Omega^{\mu\nu}+\left(5R^2+2R_{\mu\nu\rho\sigma}R^{\mu\nu\rho\sigma}-2R_{\mu\nu}R^{\mu\nu} \right)I\right\rbrace.
\end{gathered}
\end{equation}

We note that all the 
total derivative terms are neglected as
they will serve as boundary terms in the integral
\cref{sd} and hence will vanish after integration. 
Here, \enquote{tr} denotes trace over the field index $m$. 
One can evaluate curvature tensors and curvature scalar 
from the background metric ($g_{\mu\nu}$) and  the results for $E$, $I$ and
$\Omega$ can be obtained from \cref{geb,omega}. In principle, we can compute
higher order Seeley-DeWitt coefficients, $a_6(x), a_8(x),\cdots$ as per requirements in this method.

\subsection{Elementary examples}\label{spinor}
In this section, we start with illustrating the general heat kernel approach to calculate the heat kernel
coefficients for Dirac spinor field as an elementary example. We calculate first three Seeley-DeWitt coefficients for 
four different cases namely massless free Dirac spinor field, massive free Dirac spinor field,
massless gauged Dirac spinor field, and massive gauged Dirac spinor field.

\subsubsection{Free Dirac spinor field}\label{free spinor}
 The classical Lagrangian for a \textit{massless} free Dirac spinor $\psi$ is given by
 \begin{equation}\label{lag1}
 \mathcal{L} = i\bar{\psi}\gamma^\mu D_\mu\psi,
 \end{equation}
 where $D_\mu$ is the covariant derivative acting on the spinors
 and the gamma matrices ($\gamma^\mu$) satisfy the standard 4D Clifford algebra,
\begin{equation}\label{gamma}
 \gamma^\mu\gamma^\nu+\gamma^\nu\gamma^\mu = 2g^{\mu\nu}\mathbb{I}_4,
\end{equation}

with $\mathbb{I}_4$ being the identity matrix in Clifford algebra. The Lagrangian \eqref{lag1} contains first-order Dirac-type  differential operator $\slashed{D} = i\gamma^\mu D_\mu$, but our aim is to make it quadratic in order.
Assuming our spacetime to be even-dimensional having Euclidean signature, our gamma matrices are Hermitian,
$\gamma_\mu^\dagger=\gamma_\mu$ and hence the operator $\gamma^\mu D_\mu$ is anti-Hermitian, 
$(\gamma^\mu D_\mu)^\dagger=-\gamma^\mu D_\mu$. With this choice, the differential operator $\slashed{D}$ is Hermitian.

We then compute the relevant second-order Laplace-type differential operator\footnote{With our 
choice of four-dimensional 
space-time, $\slashed{D}$ is Hermitian and we can express $\Lambda=\slashed{D}^\dagger\slashed{D}=\slashed{D}^2$.} 
acting on $\psi$,
\begin{equation} \label{laplace1}
\begin{split}
\Lambda = \slashed{D}^2 &= i^2\gamma^\mu\gamma^\nu D_\mu D_\nu\\
&=-(D^\mu D_\mu)-\frac{1}{2}\gamma^\mu\gamma^\nu[D_\mu,D_\nu].
\end{split}
\end{equation}

Gamma matrix commutation relation gives,
\begin{equation}\label{def1}
 [D_\mu,D_\nu]\psi = \frac{1}{4}\gamma^\theta\gamma^\phi R_{\mu\nu\theta\phi}\psi.
\end{equation}
Also, using  gamma identity \eqref{gamma} and standard Riemann identities, we can express
\begin{equation}\label{spec}
\thinspace \gamma^\mu\gamma^\nu\gamma^\theta\gamma^\phi R_{\mu\nu\theta\phi} = -2R.
\end{equation}
Inserting relations \eqref{def1} and \eqref{spec} in eq.\eqref{laplace1}, we find that
\begin{equation} \label{laplace2}
\bar{\psi} \Lambda\psi =
-\bar{\psi}(D^\mu D_\mu)\psi+\frac{R}{4}\bar{\psi}\psi.
\end{equation}
Finally, we can rewrite  eq.\eqref{laplace2} using the definition $\mathcal{D}_\mu = D_\mu + \omega_\mu$,
  \begin{equation} \label{laplace3}
  \begin{split}
   \bar{\psi} \Lambda\psi &=
   \bar{\psi}\left\lbrace -\mathcal{D}^\mu\mathcal{D}_\mu+ D^\mu\omega_\mu+\omega_\mu\omega^\mu+
\frac{R}{4} + 2\omega^\mu D_\mu\right\rbrace\psi.
  \end{split}
   \end{equation}

The form of $\Lambda$ given in \cref{laplace3} is of the Laplace-type form as required in \cref{compp} and one can compare this $\Lambda$ with \cref{comp} and easily extract following useful matrices with the help of \cref{geb}
\begin{equation}\label{s2}
\begin{gathered}
 I = \mathbb{I}_4, \enspace \omega_\mu = 0,\enspace E = -\frac{R}{4}\mathbb{I}_4,\enspace \Omega_{\mu\nu} = \frac{1}{4}\gamma^\theta\gamma^\phi R_{\mu\nu\theta\phi},
\end{gathered}
\end{equation}
where in calculating $ \Omega_{\mu\nu}$ we have used \cref{def1} in the definition of $\Omega$ 
 given in \cref{omega}.

Next, we will use gamma contraction identities \cite{Pal:2007zx} and compute all the necessary traces which are useful for obtaining Seeley-DeWitt coefficients,
\begin{equation}\label{x1}
\text{tr}(I)=4,\enspace \text{tr}(E)=-R,\enspace \text{tr}(E^2)=\frac{R^2}{4}, \enspace \text{tr}(\Omega_{\mu\nu}\Omega^{\mu\nu})= -\frac{1}{2}R_{\mu\nu\rho\sigma}R^{\mu\nu\rho\sigma}.
\end{equation}
Inserting these data into \cref{coefficients}, we find Seeley-DeWitt coefficients for \textit{massless} free Dirac spinor field as
\begin{equation}
 \begin{gathered}
(4\pi)^2a_0(x) = -4,\\
(4\pi)^2a_2(x) = +\frac{R}{3},\\
(4\pi)^2a_4(x) = -\frac{1}{360}(5R^2-8R_{\mu\nu}R^{\mu\nu}-7R_{\mu\nu\rho\sigma}R^{\mu\nu\rho\sigma}).
\end{gathered}
\end{equation}
Note that we have put an overall minus sign by hand on 
each Seeley-DeWitt coefficients to account for fermion statistics \cite{Peixoto:2001wx} and we will do the same for all the spinor cases.

Our next aim is to study the case of \textit{massive} free Dirac spinor (with real mass $m$) and  for that, we have to add a mass term in the Lagrangian given in \cref{lag1},
 \begin{equation}
 \mathcal{L} = \bar{\psi}\left\lbrace i\gamma^\mu D_\mu+m\right\rbrace\psi,
 \end{equation}
and the corresponding Laplace-type operator can be obtained by following the technique 
discussed in \cite{Peixoto:2001wx},
 \begin{equation}
 \begin{split}
\Lambda &= \left( i\gamma^\mu D_\mu+m\right)\times \left( i\gamma^\nu D_\nu-m\right)\\
 &= -\gamma^\mu\gamma^\nu D_\mu D_\nu-m^2.
 \end{split}
 \end{equation}
Proceeding in the same way as for massless Dirac spinor field, we obtain
\begin{equation}
\begin{gathered}
E = (m^2-\frac{R}{4})\mathbb{I}_4,\\
\text{tr}(E) = 4m^2-R,\\
\enspace \text{tr}(E^2) = 4m^4-2m^2R+\frac{R^2}{4},
\end{gathered}
\end{equation}
while other matrices like $I, \omega, \Omega$ and their traces remain exactly same as in \cref{s2,x1}.

Using the results in \cref{coefficients}, we obtain the Seeley-DeWitt coefficients for \textit{massive} free Dirac spinor field as
\begin{equation}\label{abcd}
 \begin{gathered}
(4\pi)^2a_0(x) = -4,\\
(4\pi)^2a_2(x) = -(4m^2-\frac{R}{3}),\\
(4\pi)^2a_4(x) = -\frac{1}{360}(720m^4-120m^2R+5R^2-8R_{\mu\nu}R^{\mu\nu}-7R_{\mu\nu\rho\sigma}R^{\mu\nu\rho\sigma}).
\end{gathered}
\end{equation}

\pagebreak
\subsubsection{ Dirac spinor field with gauge connection}\label{gauge spinor}
The standard form of Lagrangian for a \textit{massless} Dirac spinor field $\psi$ with gauge field  connection $A_\mu$ is given by
\begin{equation}
 \mathcal{L} = i\bar{\psi}\gamma^\mu (D_\mu+A_\mu)\psi.
\end{equation}
With our choice of space-time which is
even dimensional and has euclidean signature,
 $A_\mu$ is anti-Hermitian in the gauge indices, $A_\mu^\dagger=-A_\mu$; also, $\gamma^\mu D_\mu$ is anti-Hermitian, $(\gamma^\mu D_\mu)^\dagger=-\gamma^\mu D_\mu$; and hence, $\slashed{D}=i\gamma^\mu(D_\mu+A_\mu)$ is Hermitian. 
We employ the same approach as we did in \cref{free spinor} and find that the relevant second-order differential operator acting on $\psi$ takes the following form
\begin{equation}\label{s3}
\begin{split}
\Lambda =\slashed{D}^2 &= i\gamma^\mu(D_\mu+A_\mu)\times i\gamma^\nu(D_\nu+A_\nu)\\
&= -D^\mu D_\mu-\frac{1}{2}\gamma^\mu\gamma^\nu[D_\mu,D_\nu]-\frac{1}{2}\gamma^\mu\gamma^\nu F_{\mu\nu}-D^\mu A_\mu-2A^\mu D_\mu-A^\mu A_\mu,
\end{split}
\end{equation}
where the notation $F_{\mu\nu}= \partial_\mu A_\nu-\partial_\nu A_\mu+[A_\mu,A_\nu]$. 

Now, using the result of \cref{laplace3} we find for massless gauged spinor field,
\begin{equation}\label{s4}
\begin{split}
\bar{\psi} \Lambda\psi
&= \bar{\psi}\left\lbrace-\mathcal{D}^\mu\mathcal{D}_\mu+ D^\mu\omega_\mu+\omega^\mu\omega_\mu-\frac{1}{2}\gamma^\mu\gamma^\nu F_{\mu\nu}-D^\mu A_\mu - A^\mu A_\mu+\frac{R}{4}+2(\omega^\mu-A^\mu)D_\mu\right\rbrace\bar{\psi},
\end{split}
\end{equation}

Finally, comparing the form of $\Lambda$ from \cref{s4} with \cref{comp} and using \cref{geb}, we identify $I$, $\omega_\mu$, $E$, and $\Omega_{\mu\nu}$ as
\begin{equation}\label{f1}
\begin{split}
 I &= \mathbb{I}_4,\enspace \omega_\mu = A_\mu,\\
E &= -D^\mu\omega_\mu-\omega_\mu\omega^\mu+\frac{1}{2}\gamma^\mu\gamma^\nu F_{\mu\nu}+D^\mu A_\mu + A^\mu A_\mu-\frac{R}{4}\mathbb{I}_4\\ &= \frac{1}{2}\gamma^\mu\gamma^\nu F_{\mu\nu}-\frac{R}{4}\mathbb{I}_4,\\
\Omega_{\mu\nu} &= [D_\mu,D_\nu]+D_{[\mu}\omega_{\nu]}+[\omega_\mu,\omega_\nu]\\
&= \frac{1}{4}\gamma^\theta\gamma^\phi R_{\mu\nu\theta\phi}+ F_{\mu\nu},
\end{split}
\end{equation}

Note that, for obtaining $\Omega$, we have substituted the relation, \eqref{def1} in the definition \eqref{omega}.

Now, our task is to calculate all the traces necessary for calculating Seeley-DeWitt coefficients. The results are
\begin{equation}\label{f2}
\begin{split}
\text{tr}(I) &= 4,\\
\text{tr}(E) &= -R,\\
\text{tr}(E^2) &= \frac{R^2}{4}-2F_{\mu\nu}F^{\mu\nu},\\
\text{tr}(\Omega_{\mu\nu}\Omega^{\mu\nu}) &= 4F_{\mu\nu}F^{\mu\nu}-\frac{1}{2}R_{\mu\nu\rho\sigma}R^{\mu\nu\rho\sigma}.
\end{split}
\end{equation}
Here, we have used standard gamma contraction identities \cite{Pal:2007zx} and Riemann identities for the necessary simplifications.

We use the result of these traces in the formulae \eqref{coefficients} to calculate the Seeley-DeWitt coefficients for the \textit{massless} gauged Dirac spinor field,
\begin{equation}
\begin{gathered}
(4\pi)^2a_0(x) = -4,\\
(4\pi)^2a_2(x) = +\frac{R}{3},\\
(4\pi)^2a_4(x) = -\frac{1}{360}(5R^2-8R_{\mu\nu}R^{\mu\nu}-7R_{\mu\nu\rho\sigma}R^{\mu\nu\rho\sigma}-240F_{\mu\nu}F^{\mu\nu}).
\end{gathered}
\end{equation}

Next, we are going to study the case of \textit{massive} gauged Dirac spinor field having the following form of Lagrangian
\begin{equation}
 \mathcal{L} = \bar{\psi}\left\lbrace i\gamma^\mu \left(D_\mu+A_\mu\right)+ m\right\rbrace\psi.
\end{equation}
To deal with the mass term, we use the same technique as we did for massive free Dirac spinor in \cref{free spinor}  and find the required second-order differential operator as
\begin{equation}
 \begin{split}
\Lambda &= \left\lbrace i\gamma^\mu (D_\mu+A_\mu)+m\right\rbrace\times \left\lbrace i\gamma^\nu (D_\nu+A_\nu)-m\right\rbrace\\
 &= -\gamma^\mu\gamma^\nu (D_\mu+A_\mu)( D_\nu+A_\nu)-m^2.
 \end{split}
 \end{equation}
Using the results from \cref{s3,s4}, we obtain the necessary matrices and traces as
 \begin{equation}
\begin{split}
 E &= \frac{1}{2}\gamma^\mu\gamma^\nu F_{\mu\nu}+(m^2-\frac{R}{4})\mathbb{I}_4,\\
\text{tr}(E) &=4m^2-R,\\
\text{tr}(E^2) &= 4m^4-2m^2R+\frac{R^2}{4}-2F_{\mu\nu}F^{\mu\nu},
\end{split}
\end{equation}
while other matrices $I, \omega$ and $\Omega$, and their traces remain exactly the same as in \cref{f1,f2}. 

Now, substituting this data in the formulae \eqref{coefficients}, we obtain Seeley-DeWitt coefficients for the \textit{massive} gauged Dirac spinor field as
\begin{equation}\label{abc}
\begin{gathered}
(4\pi)^2a_0(x) = -4,\\
(4\pi)^2a_2(x) = -(4m^2-\frac{R}{3}),\\
(4\pi)^2a_4(x) = -\frac{1}{360}(720m^4-120m^2R+5R^2-8R_{\mu\nu}R^{\mu\nu}-7R_{\mu\nu\rho\sigma}R^{\mu\nu\rho\sigma}-240F_{\mu\nu}F^{\mu\nu}).
\end{gathered}
\end{equation}

\section{Spin-3/2 Rarita-Schwinger field}\label{sec3}
The spin-3/2 particle theory was first proposed by W. Rarita and J. Schwinger \cite{Rarita:1941wz}, 
based on Fierz and Pauli construction of field theory with arbitrary spins.
In four-dimensional space-time, the Lagrangian of a massless free RS field 
$\psi_{\mu }$ is given by 
\begin{equation}
\mathcal{L}_{RS}= -i\bar{\psi}_{\mu }\gamma^{\mu\lambda\nu}D_\lambda\psi_{\nu },
\end{equation}
where $\gamma^{\mu\lambda\nu}$ is the
antisymmetrized product of gamma matrices with 
respect to the indices $\mu$, $\lambda$, and $\nu$. 
Using the relation \eqref{gamma}, 
one can simplify $\gamma^{\mu\lambda\nu}$ and 
rewrite $\mathcal{L}_{RS}$ as
\begin{equation}\label{RS}
 \mathcal{L}_{RS}= -\frac{i}{2}\bar{\psi}_{\mu }(\gamma^\mu\gamma^\lambda\gamma^\nu-\gamma^\nu\gamma^\lambda\gamma^\mu)D_\lambda\psi_{\nu }.
\end{equation}
We gauge fix the theory by adding the following gauge fixing term to the Lagrangian in \cref{RS}
\begin{equation}
\mathcal{L}_{GF} = \frac{i}{2}\bar{\psi}_{\mu }\gamma^\mu\gamma^\lambda\gamma^\nu D_\lambda\psi_{\nu }.
\end{equation}
As a result, the total gauge fixed Lagrangian takes the following form
\begin{equation}
\mathcal{L} = \mathcal{L}_{RS}+\mathcal{L}_{GF} = \frac{i}{2}\bar{\psi}_{\mu }\gamma^\nu\gamma^\lambda\gamma^\mu D_\lambda\psi_{\nu }.
\end{equation}
The Lagrangian consists of a first-order 
Dirac-type differential operator
$\slashed{D}^{\mu\nu} \equiv \frac{i}{2}\gamma^\nu\gamma^\lambda\gamma^\mu D_\lambda$ and
with our choice of space-time, $\slashed{D}$ is Hermitian. 
In order to obtain the corresponding second-order Laplace-type operator, let's consider
\begin{equation} \label{rel1}
\begin{split}
 \Lambda^{\mu\nu} = {(\slashed{D}^2)}^{\mu\nu} &= \slashed{D}^{\mu\sigma}{\slashed{D}_\sigma}^\nu \\
&= \left( \frac{i}{2}\gamma^\sigma\gamma^\rho\gamma^\mu D_\rho\right)\times\left( \frac{i}{2}\gamma^\nu\gamma^\lambda\gamma_\sigma D_\lambda\right)\\
&= -\frac{1}{2}(\gamma^\lambda\gamma^\rho\gamma^\mu\gamma^\nu+\gamma^\nu\gamma^\mu\gamma^\rho\gamma^\lambda )D_\rho D_\lambda\\
&= -g^{\mu\nu}(D^\rho D_\rho)+\frac{1}{4}
( \gamma^\rho\gamma^\lambda\gamma^\mu\gamma^\nu+\gamma^\nu\gamma^\mu\gamma^\lambda\gamma^\rho)[D_\rho, D_\lambda],
\end{split}
\end{equation}
where equality in the third line is obtained
using the gamma matrix contraction identity \cite{Pal:2007zx}.\\
In order to simplify the expression \cref{rel1}, we now consider
\begin{equation}
\label{lap1}
\begin{split}
\bar{\psi}_{\mu }\Lambda^{\mu\nu}\psi_{\nu }
&= \bar{\psi}_{\mu }\left\lbrace-g^{\mu\nu}D^\rho D_\rho+\frac{1}{4}( \gamma^\rho\gamma^\lambda\gamma^\mu\gamma^\nu+\gamma^\nu\gamma^\mu\gamma^\lambda\gamma^\rho)[D_\rho, D_\lambda]\right\rbrace\psi_{\nu }\\
\end{split}
\end{equation}

Now, one can proceed in a systematic manner and simplify \cref{lap1} using the following identities
\begin{equation}\label{cal}
\begin{gathered}
[D_\rho,D_\lambda]\psi_{\nu } = 
-{R^\sigma}_{\nu\rho\lambda}\psi_{\sigma }+\frac{1}{4}R_{\rho\lambda\theta\phi}\gamma^\theta\gamma^\phi\psi_{\nu },\\
(\gamma^\rho\gamma^\lambda\gamma^\mu\gamma^\sigma+\gamma^\sigma\gamma^\mu\gamma^\lambda\gamma^\rho){R^\nu}_{\sigma\rho\lambda}=-4R^{\mu\nu},\\
( \gamma^\rho\gamma^\lambda\gamma^\mu\gamma^\nu+\gamma^\nu\gamma^\mu\gamma^\lambda\gamma^\rho)R_{\rho\lambda\theta\phi}\gamma^\theta\gamma^\phi=8(-\gamma^\rho\gamma^\theta {R^{\mu\nu}}_{\rho\theta}+\gamma^\mu\gamma^\theta{R^\nu}_\theta-\gamma^\nu\gamma^\theta{R^\mu}_\theta)\\
\hspace{3.5in}-4R\left(\gamma^\mu\gamma^\nu-g^{\mu\nu} \right),
\end{gathered}
\end{equation}
to get the final form of Laplace-type differential operator as
\begin{equation}\label{final}
\begin{gathered}
\Lambda^{\mu\nu} =  -g^{\mu\nu}\mathcal{D}^\rho \mathcal{D}_\rho+R^{\mu\nu}-\frac{1}{2}\gamma^\rho\gamma^\theta {R^{\mu\nu}}_{\rho\theta}+\frac{1}{2}\gamma^\mu\gamma^\theta{R^\nu}_\theta-\frac{1}{2}\gamma^\nu\gamma^\theta{R^\mu}_\theta-\frac{R}{4}\gamma^\mu\gamma^\nu+\frac{R}{4}g^{\mu\nu}.
\end{gathered}
\end{equation}
Note that, we  have set $\omega = 0$ in the following definition as $\omega$ 
vanishes for any free field,
\begin{equation}\label{def4}
\mathcal{D}_\alpha\psi_{\mu } = D_\alpha\psi_{\mu } + {(\omega_\alpha)_{\mu }}^{\nu }\psi_{\nu }.
\end{equation}
Comparing $\Lambda$ from \cref{final} with \cref{comp}, we identify  $I$, $E$ and $\Omega_{\mu\nu}$ as
\begin{equation}\label{E}
\begin{split}
 I^{\mu  \nu } &= \mathbb{I}_4 g^{\mu\nu},\\
E^{\mu  \nu } & = -\mathbb{I}_4 R^{\mu\nu}+\frac{1}{2}\gamma^\rho\gamma^\theta {R^{\mu\nu}}_{\rho\theta}-\frac{1}{2}\gamma^\mu\gamma^\theta{R^\nu}_\theta +\frac{1}{2}\gamma^\nu\gamma^\theta{R^\mu}_\theta+\frac{R}{4}\gamma^\mu\gamma^\nu-\frac{R}{4}\mathbb{I}_4 g^{\mu\nu},\\
(\Omega_{\mu\nu})^{\alpha \beta }  &=\mathbb{I}_4 {R^{\alpha\beta}}_{\mu\nu}+\frac{1}{4}g^{\alpha\beta}R_{\mu\nu\rho\lambda}\gamma^\rho\gamma^\lambda,
\end{split}
\end{equation}
where we have used \cref{cal,def4} in the definition of $\Omega$ given in \cref{omega} to obtain its present form.
\subsection{Trace calculations}\label{trace}
Now, we compute the traces necessary 
for evaluation of Seeley-DeWitt coefficients using the matrices \eqref{E}. 
Note that for this purpose, we will make use of Riemann identities,
and the necessary gamma trace and contraction identities \cite{Pal:2007zx}. The trace results are as follows:\\

\begin{equation}
\text{tr}(I) = 16.
\end{equation} 
\begin{equation}
\begin{split}
\text{tr}(E) &=g_{\mu\nu} \left\lbrace -\mathbb{I}_4 R^{\mu\nu}+\frac{1}{2}\gamma^\rho\gamma^\theta {R^{\mu\nu}}_{\rho\theta}-\frac{1}{2}\gamma^\mu\gamma^\theta{R^\nu}_\theta +\frac{1}{2}\gamma^\nu\gamma^\theta{R^\mu}_\theta+\frac{R}{4}\gamma^\mu\gamma^\nu-\frac{R}{4}\mathbb{I}_4 g^{\mu\nu}\right\rbrace\\
&=-4R,
\end{split}
\end{equation}
\begin{equation}\label{t1}
\begin{split}
\text{tr}(E^2) &= \left( \underbrace{-\mathbb{I}_4 R^{\mu\nu}}_{A}\overbrace{+\frac{1}{2}\gamma^\rho\gamma^\theta {R^{\mu\nu}}_{\rho\theta}}^{B}\underbrace{-\frac{1}{2}\gamma^\mu\gamma^\theta{R^\nu}_\theta}_{C}\overbrace{+\frac{1}{2}\gamma^\nu\gamma^\theta{R^\mu}_\theta}^{D}\underbrace{+\frac{R}{4}\gamma^\mu\gamma^\nu}_{F}\overbrace{-\frac{R}{4}\mathbb{I}_4 g^{\mu\nu}}^{G}\right)\\
&\enspace\times \left( \overbrace{-\mathbb{I}_4 R_{\nu\mu}}^{1}\underbrace{+\frac{1}{2}\gamma^\sigma\gamma^\phi R_{\nu\mu\sigma\phi}}_{2}\overbrace{-\frac{1}{2}\gamma_\nu\gamma^\phi R_{\mu\phi}}^{3}+\underbrace{\frac{1}{2}\gamma_\mu\gamma^\phi R_{\nu\phi}}_{4}\overbrace{+\frac{R}{4}\gamma_\nu\gamma_\mu}^{5}\underbrace{-\frac{R}{4}\mathbb{I}_4 g_{\nu\mu}}_{6}\right).
\end{split}
\end{equation}
Note that each term in \cref{t1} is denoted by a corresponding symbol, and now we are going compute their products separately as follows
\begin{equation}\nonumber
\begin{split}
A1 &= 4 R_{\mu\nu}R^{\mu\nu}\\
B2 &= 2R_{\mu\nu\rho\sigma}R^{\mu\nu\rho\sigma}\\
C3 &=R^2\\
D4 &= R^2\\
F5 &= 4R^2\\
G6 &= R^2
\end{split}
\hspace{0.5in}
\begin{split}
A2 =B1 &= 0\\
A3 = C1 &= 2R_{\mu\nu}R^{\mu\nu}\\
A4 =D1 &= -2R_{\mu\nu}R^{\mu\nu}\\
A5 = F1 &= -R^2\\
A6 = G1 &= R^2\\
B3 =C2 &= -2R_{\mu\nu}R^{\mu\nu}\\
B4 = D2 &= -2R_{\mu\nu}R^{\mu\nu}\\
B5 = F2 &= R^2
\end{split}
\hspace{0.5in}
\begin{split}
B6 = G2 & = 0\\
C4 = D3 &= 2R_{\mu\nu}R^{\mu\nu}\\
C5 = F3 &= -2R^2\\
C6= G3 &= \frac{1}{2}R^2\\
D5 = F4 &= -R^2\\
D6 = G4 &= -\frac{1}{2}R^2\\
F6 = G5 &= -R^2
\end{split}
\end{equation}
One may use the results of these terms in \cref{t1} and finally can come up with the following result
\begin{equation}
\text{tr}(E^2)=2R_{\mu\nu\rho\sigma}R^{\mu\nu\rho\sigma}+R^2.
\end{equation} 
We will also use the result for $\Omega$ from \cref{E} and evaluate
\begin{equation}
\begin{split}
\text{tr}(\Omega_{\mu\nu}\Omega^{\mu\nu}) &= \left\lbrace \mathbb{I}_4 {R^{\alpha\beta}}_{\mu\nu}+\frac{1}{4}g^{\alpha\beta}\gamma^\rho\gamma^\lambda R_{\mu\nu\rho\lambda}\right\rbrace \times \left\lbrace \mathbb{I}_4 {R_{\beta\alpha}}^{\mu\nu}+\frac{1}{4}g_{\beta\alpha}\gamma^\theta\gamma^\phi{R_{\mu\nu}}_{\theta\phi}\right\rbrace\\
&= -6R_{\mu\nu\rho\sigma}R^{\mu\nu\rho\sigma}.
\end{split}
\end{equation}
\subsection{Seeley-DeWitt coefficients}
One can substitute all the traces calculated in \cref{trace} in the standard formulae \eqref{coefficients}
in order to find the leading Seeley-DeWitt coefficients for the case of
massless free spin-3/2 Rarita-Schwinger fermions. The results are
\begin{equation}\label{xyz1}
\begin{gathered}
(4\pi)^2a_0(x) = -16,\\
(4\pi)^2a_2(x) = +\frac{4}{3}R,\\
(4\pi)^2a_4(x) = -\frac{1}{360}(212R_{\mu\nu\rho\sigma}R^{\mu\nu\rho\sigma}-32R_{\mu\nu}R^{\mu\nu}+20R^2).
\end{gathered}
\end{equation}
For the background satisfying the condition $R=0$, one has 
to ignore the terms proportional to R in \cref{xyz1}. In case of Einstein manifold, 
one can arrive at the following results by setting $R_{\mu\nu}=\lambda g_{\mu\nu}$ 
($\lambda$ is arbitrary constant) in \cref{xyz1}
\begin{equation}\label{xyz2}
\begin{gathered}
(4\pi)^2a_0(x) = -16,\\
(4\pi)^2a_2(x) = +\frac{16}{3}\lambda,\\
(4\pi)^2a_4(x) = -\frac{1}{360}(212R_{\mu\nu\rho\sigma}R^{\mu\nu\rho\sigma}+192\lambda^2).
\end{gathered}
\end{equation}
\end{ceqn}
Setting $\lambda=0$ in \cref{xyz2}, will provide the results for the Ricci-flat space-time ($R_{\mu\nu}=0$).
Note that, here also we have placed an overall minus sign by hand on each coefficient to account for fermion statistics. 


The above 
results are clearly consistent with ref.\cite{Charles:2015nn} for $R=0$ manifold,
where the authors have calculated the Seeley-DeWitt
coefficients for fermions in  \textit{$\mathcal{N}=2$} gravity multiplet\footnote{see section 3.4.3
of ref.\cite{Charles:2015nn}}. One can extract the results for single free gravitino field, by 
ignoring 
the field-type indices $A, B$ and setting $F_{\mu\nu}=0$ in eqs 3.79 and 3.80 of ref.\cite{Charles:2015nn}.
Also, for the special case of  Ricci-flat manifold one can check the consistency from 
ref.\cite{Endo:1995wz}. Considering the result for $d = 4$  and performing  appropriate 
 traces over eq. 24 of ref.\cite{Endo:1995wz}, 
 one can obtain the Seeley-DeWitt coefficients which are identical with our result \eqref{xyz2} for $\lambda =0$.

\section{Summary}\label{diss}
In this report, we have computed the first three Seeley-DeWitt coefficients for the massless free 
spin-3/2 field and expressed them in terms of the local invariants of the theory.  We started with the discussion 
of the general heat kernel approach and worked out
a practice calculation for Seeley-DeWitt coefficients for spin-1/2 field as an elementary example. 
Then, we applied  the same approach to calculate the Seeley-DeWitt coefficients for 
spin-3/2 Rarita-Schwinger field and discussed about the consistency of our results for various special cases. 
Apart from various interests, spin-3/2 Rarita-Schwinger field has a particular
role in supergravity and evolution of the universe. However,
we are interested in computing the heat kernel coefficients for the same in 
any arbitrary background as these coefficients have various applications in quantum gravity,
quantum field theory, and black hole entropy corrections (mainly logarithmic corrections). 
One can employ these results of Seeley-DeWitt coefficients to 
compute the heat kernel coefficients by 
integrating them over the manifold of interest as discussed in 
\cref{formulae}. 
We also note that the derivation in this paper assumes that $\psi$ is a Dirac spinor. Anyone
interested in 
computing the heat kernel coefficients for  Majorana or Weyl fermions instead of Dirac fermions
can simply divide the results by two, since Weyl and Majorana spinors have half the degrees of freedom 
of Dirac spinor. 

\section{Acknowledgments}
We would like to thank Gourav Banerjee for useful discussions. B.P. acknowledges IIT(ISM), Dhanbad for the Grant (FRS (53)/2013-2014/APH).


\end{document}